Title: Multi-scale community organization of the human structural connectome and its relationship with resting-state functional connectivity

Authors: Richard F. Betzel[1,2], Alessandra Griffa[3,4], Andrea Avena-Koenigsberger[1,2], Joaquín Goñi[1], Jean-Phillippe Thiran[3,4], Patric Hagmann[3,4], Olaf Sporns[1,2].

Affiliations:
[1]Department of Psychological and Brain Sciences – Indiana University, Bloomington, USA
[2]Program in Cognitive Science – Indiana University, Bloomington, USA
[3]Department of Radiology, University Hospital Center and University of Lausanne (CHUV) – Lausanne, Switzerland
[4]Signal Processing Laboratory (LTS5), Ecole Polytechnique Fédérale de Lausanne (EPFL) – Lausanne, Switzerland

Abstract – 176 words
Body – 6,135 words
Figures – 8
Tables – None
Supplement – community_assignments.txt, names.txt, roi.txt

**Abstract**

The human connectome has been widely studied over the past decade. A principal finding is that it can be decomposed into communities of densely interconnected brain regions. This result, however, may be limited methodologically. Past studies have often used a flawed modularity measure in order to infer the connectome's community structure. Also, these studies relied on the intuition that community structure is best defined in terms of a network's static topology as opposed to a more dynamical definition. In this report we used the partition stability framework, which defines communities in terms of a Markov process (random walk), to infer the connectome's multi-scale community structure. Comparing the community structure to observed resting-state functional connectivity revealed communities across a broad range of dynamical scales that were closely related to functional connectivity. This result suggests a mapping between communities in structural networks, models of communication processes, and brain function. It further suggests that communication in the brain is not limited to a single characteristic scale, leading us to posit a heuristic for scale-selective communication in the cerebral cortex.

**Introduction**

Many complex networks exhibit community structure, defined for example by clustered edge distributions such that vertices (nodes) in the same community preferentially link to one another (Guimera & Amaral, 2005; Girvan & Newman, 2004; Newman, 2006). Examples of community structure can be found in society as groups of friends, workplaces, cities, and states (Moody & White, 2003; Freeman, 2004); in protein interaction networks as groups of co-functioning proteins (Jonsson et al., 2006); and in the World Wide Web (WWW) as webpages sharing many hyperlinks (Albert et al., 1999; Flake et al., 2002).

Detecting community structure is an important endeavor in network science. Though there exist many methods for doing so, none has emerged as clearly preeminent (for a review, see Fortunato, 2010). One of the most widely used methods is to identify the vertex partition that maximizes the modularity quality function, defined as the difference between observed and expected intra-community edge density (Newman & Girvan, 2004). Despite its widespread usage, modularity suffers from several drawbacks: maximization is NP-hard and identifying the optimal partition is often practically impossible (Fortunato, 2010); solutions are sometimes degenerate, with multiple partitions corresponding to the maximum modularity (Good et al., 2010); and a resolution limit renders modularity "blind" to communities below some characteristic scale (Fortunato & Barthelemy, 2007).

The first two problems are ubiquitous to most community detection algorithms and are generally unavoidable. The resolution limit is more specific to quality functions like modularity, but has been remedied in several instances by the inclusion of a tunable resolution parameter, which controls the scale at which communities are detected (Reichardt & Bornholdt, 2006; Arenas et al., 2008; Ronhovde & Nussinov, 2009). One such example is the partition stability framework, which defines communities in terms of a Markov process based on a random walk model (Lambiotte et al, 2008; Delvenne et al., 2010; Lambiotte, 2010). As this process evolves, a random walker makes progressively longer walks and explores more distant parts of the network. Intuitively, communities can be thought of as groups of vertices that effectively "trap" the flow of random walkers over a particular dynamical scale. In this way, the stability framework has been dubbed a "zooming lens," whereby focusing in on shorter or longer dynamical timescales reveals communities of correspondingly smaller or larger diameter (Schaub et al., 2012).

The human connectome, i.e. the full set of neural elements and connections of the human brain, can be modeled as a complex network (Sporns et al., 2005; Bullmore & Sporns, 2009). The topological properties of this network have been studied for nearly a decade, revealing key features including small-world architecture (Gong et al., 2009), hub regions and cores (Hagmann et al., 2008), rich club organization (van den Heuvel & Sporns, 2011), modular architecture (Chen et al., 2008; Meunier et al., 2010; Wu et al., 2011), and economical wiring (Bassett et al., 2010; Bullmore & Sporns, 2012). While these results characterize and contextualize the connectome among all complex networks, the role of these topological features in shaping communication processes and dynamic couplings among brain regions remains an area of active research (Honey et al., 2009; van den Heuvel et al., 2012, Haimovici et al., 2013).

Empirical and computational studies suggest that the human connectome underpins complex neural dynamics and facilitates the communication and integration of information between brain regions. In functional magnetic resonance imaging (fMRI), such "functional connectivity" is reflected by the magnitude of statistical dependence, commonly measured as linear correlation, between blood oxygen-level dependent (BOLD) signals recorded from different brain regions. A growing body of literature describes patterns of resting-state functional connectivity (rsFC), i.e. spontaneous, endogenous fluctuations of the BOLD signal in the absence of any explicit cognitive task. This work has revealed consistent connectivity patterns, dubbed resting-state networks (RSN), which resemble networks of regions (e.g. somato-motor, visual, default mode, etc.) that are coherently engaged in various cognitive and behavioral domains (Greicius et al., 2003; Damoiseaux et al., 2006; Fox & Raichle, 2007, Smith et al., 2009).

This article has two principal aims. The first aim consists of detecting and characterizing the multi-scale community structure of the human connectome using the partition stability framework. The second aim is to assess the relationship of such structure to empirically observed rsFC. Where the first aim attempts to answer the question 'what communities of brain regions are poised for diffusion-like communication', the second aim attempts to answer the question 'what evidence is there that the previously-identified communities match observed patterns of functional coupling in the brain and at which dynamical scales is the correspondence between community structure and functional coupling most salient?'

**Methods**

Forty (40) healthy human volunteers (24 males and 16 females, 25.3±4.9 years old) underwent an MRI session on a 3-Tesla scanner (Trio, Siemens Medical, Germany) with a 32-channel head-coil. The session consisted of (i) a magnetization-prepared rapid acquisition gradient echo (MPRAGE) sequence sensitive to white/grey matter contrast (1 mm in-plane resolution, 1.2 mm slice thickness.), (ii) a diffusion spectrum imaging (DSI) sequence (128 diffusion weighted volumes + 1 reference $b_0$ volume, maximum $b$-value 8000 s/mm$^2$, 2.2 × 2.2 × 3.0 mm voxel size), and (iii) a gradient echo EPI sequence sensitive to BOLD contrast (3.3 mm in-plane resolution and slice thickness with a 0.3 mm gap, TR 1920 ms). During the fMRI acquisition, subjects were lying in the scanner with eyes open, resting but awake and cognitively alert, thus recording-resting-state fMRI (rs-fMRI).

DSI, rs-fMRI and MPRAGE data were processed using the Connectome Mapping Toolkit (Daducci et al., 2012). Each participant's gray and white matter compartments were segmented from the MPRAGE volume. In the present study the entire cortical volume was subdivided into 1000 equally-sized regions of interest (Cammoun et al., 2012). Whole brain streamline tractography was performed on reconstructed DSI data (Wedeen et al., 2008), and participant-wise, right hemisphere connectivity matrices were estimated by selecting the streamlines connecting each pair of 501 cortical regions in the right hemisphere. Connectivity strength was quantified as fiber density, as described in (Hagmann et al., 2008). Each structural connectivity (SC) matrix can be interpreted as the adjacency matrix $A_{ij}$ of a graph $G \equiv \{V, E\}$, with vertices $V = \{v_1, \dots, v_n\}$ corresponding to cortical regions of interest, and weighted, undirected edges $E = \{e_{ij}, \dots, e_{kl}\}$ representing anatomical connections. This network is often referred to as a structural connectome. It is convenient at this point to define a connectome's vertex strength and total weight as $s_i = \Sigma_j A_{ij}$ and $m = \frac{1}{2}\Sigma_{ij} A_{ij}$, respectively.

Functional data were pre-processed according to state of the art pipelines (Murphy et al., 2009; Power et al., 2012), which included motion correction, white matter, cerebrospinal fluid (CSF), global and movement signals regression, linear detrending, scrubbing and low-pass filtering. Average signals were then computed for each cortical region. The functional connectivity between each pair of regions was estimated as the Pearson correlation between the corresponding average signals. A single, right hemisphere rsFC matrix $F_{ij}$ representative of the whole group of participants was computed by averaging the 40 individual correlation matrices. The group-mean structural connectome (weighted and binary) and functional connectivity matrix for the right hemisphere of cerebral cortex are shown in Figure 1.

One of the main aims of this article was to detect the connectome's multi-scale community structure using the partition stability framework. This framework defines communities in terms of a Markov process – in this article, a continuous-time random walk. Such a process can also be used to model the diffusion of information or energy over a network, and is defined by the dynamical system:

$$\dot{p}_i = -\sum_j L_{ij} p_j$$

where $p_i$ is the probability of finding a random walker on vertex $v_i$ and $L_{ij} = \delta_{ij} - A_{ij}/s_j$ is the normalized graph Laplacian matrix. If the network is undirected and connected, this system evolves to the equilibrium state $p_i^* = s_i/2m$.

The stability framework detects community structure at different timescales of the random walk by identifying the vertex partition that maximizes the quality function "stability". Let $\mathcal{P} = \{\mathcal{C}_1, ..., \mathcal{C}_K\}$ be a partition of $V$ into $K$ communities, such that $\mathcal{C}_i \cap \mathcal{C}_j = \emptyset$ and $\cup_i \mathcal{C}_i = V$. The stability of $\mathcal{P}$ at a given Markov time $t$ is defined as:

$$R(\mathcal{P}, t) = \sum_{\mathcal{C} \in \mathcal{P}} \sum_{ij \in \mathcal{C}} \left[ (e^{-tL})_{ij} p_j^* - p_i^* p_j^* \right]$$

where the summation extends over all communities and the edges that fall within each community. The first term in the summation, $(e^{-tL})_{ij} p_j^*$, is the probability that a random walker starting in community $\mathcal{C}$ will be in that community at Markov time $t$. The second term $p_i^* p_j^*$ is the probability that two independent random walkers will be in $\mathcal{C}$ at equilibrium. The difference in these terms represents the density of random walkers in a community in excess of what is expected at equilibrium. Key to this framework, the stability measurement depends not only on the partition $\mathcal{P}$, but also on Markov time $t$. In general, different partitions will maximize stability at different times in the random walk. Varying $t$ across timescales and maximizing stability recovers the community structure at each scale.

The process of actually optimizing $R(\mathcal{P}, t)$ can be accomplished in a number of ways. One appealing option, and the one used in this article, makes use of the relationship between partition stability and a more widely used modularity measure. Lambiotte et al. (2008, 2011) demonstrated that stability could be recast in terms of the modularity of a weighted, symmetric

"flow graph." A flow graph is a transformation of $A_{ij}$ in which the dynamics of a Markov process are embedded into the edges of a new graph. In the case of a continuous-time random walk, the flow graph is represented by the full matrix $A'_{ij}(t) = (e^{-tL})_{ij} s_j$ whose elements are proportional to the probabilistic flow of random walkers between vertices at time $t$. Examples of flow graphs for one of the participants are shown in Figure 2. The relationship between stability and modularity is such that instead of directly maximizing $R(\mathcal{P}, t)$, one can equivalently maximize the modularity of a flow graph evaluated at $t$:

$$Q(\mathcal{P}, t) = \frac{1}{2m} \sum_{\mathcal{C} \in \mathcal{P}} \sum_{ij \in \mathcal{C}} \left[ A_{ij}(t) - \frac{s_i s_j}{2m} \right]$$

This result confers the practical advantage that any heuristic previously used to maximize modularity can now be repurposed in order to maximize stability.

To apply this set of principles to the connectomes obtained from 40 healthy individuals, a range of Markov times had to be selected over which to maximize stability. After experimentation, it was determined that at Markov times below $10^{-3.5}$, every participant's community structure was characterized by a partition into $n$ communities, i.e. every vertex was assigned to its own community, and at times greater than $10^{3.5}$ every participant exhibited a two-community division. Using these times as lower and upper boundaries, we selected 185 logarithmically-spaced time points over the interval $[10^{-3.5}, 10^{3.5}]$ at which to maximize partition stability. This process entailed defining a flow graph $A'_{ij}(t)$ for each participant, evaluating that flow graph at each of the pre-selected Markov times, and subsequently maximizing each flow graph's modularity (or equivalently, stability) by applying the Louvain algorithm 750 times (Blondel et al., 2008). Then, at a given time point, a participant's community structure was represented by the partition corresponding to the maximum modularity (stability).

Rather than analyze the multi-scale community structure of each participant individually, communities were aggregated across all participants to shift focus onto the group-average community structure. This operation was summarized by the weighted, symmetric and time-dependent agreement matrix $D_{ij}(t)$, whose elements indicated the percentage of all participants in which, at time $t$, assigned vertices $v_i$ and $v_j$ to the same community. Thus, the elements of $D_{ij}(t)$ were interpreted as the probability that two vertices belonged to the same community. Values ranged from '0' in the case of two vertices that never appeared in the same community, to '1' for two vertices that always appeared in the same community across all participants. Conceptually, $D_{ij}(t)$ reflected the extent to which participants' community structures at each Markov time coincided with one another.

The agreement matrix $D_{ij}(t)$ proffered a probabilistic description of the connectome's community structure at a range of Markov times. It was of practical interest to obtain at each time point a single partition corresponding to the connectome's consensus communities, i.e. the communities that were common to the majority of participants. To obtain such community structure, an iterative thresholding/clustering algorithm was applied to $D_{ij}(t)$ (Lancichinetti & Fortunato, 2012). This procedure consisted of two steps: (i) All edges in $D_{ij}(t)$ below a threshold $\tau = 0.5$ were regarded as "noise" and were set to zero. Imposing such a threshold resulted in the matrix $D'_{ij}(t)$ whose remaining edges linked only those vertices assigned to the same community greater than half of the time; (ii) $D'_{ij}(t)$ was then clustered 100 times using the Louvain algorithm. If the resulting partitions were identical, then the algorithm had reached consensus and it terminated. Otherwise, a meta-agreement matrix was built from the new partitions and the algorithm returned to (i). At every time step, then, this procedure found a consensus partition $\mathcal{P}(t)$ that captured only the features of community structure common to the majority of participants and ignored the features unique to individuals. The ordered set of consensus partitions was summarized as $\Phi = \{\mathcal{P}(t = 10^{-3.5}), \dots, \mathcal{P}(t = 10^{3.5})\}$. It should be noted that, in general, the process of obtaining a consensus partition resulted in a loss of information, removing the features of community structure that were expressed infrequently in participants. While such information might reveal meaningful differences between subjects, it was disregarded in all further analyses, which focus on characterizing features of community structure common among the cohort of participants.

A major aim of this article was to gain insight into how the connectome's multi-scale community structure related to brain function. To map this relationship, we compared the agreement matrix $D_{ij}(t)$ and the sequence of consensus partitions $\Phi$ to the empirically-measured rsFC matrix $F_{ij}$.

First, as a measure of correspondence between rsFC and community structure, we computed the Pearson correlation of the elements in the empirically-measured rsFC matrix $F_{ij}$ with $D_{ij}(t)$ at each Markov time. Larger correlation values implied that the propensity for two vertices to share a community assignment was a good predictor of whether those same vertices were functionally coupled.

As a second measure of correspondence, each consensus community's "goodness" was determined by imposing it upon the rsFC matrix $F_{ij}$. Conceptually, a "good" community was one whose internal density of functional connections was greater than expected by chance. This intuition of a community's "goodness" was in line with the way communities are defined by the modularity function. Therefore, a consensus community's "goodness" with respect to rsFC was estimated by computing its modularity score. Before doing so, we defined two additional matrices $F^+_{ij}$ and $F^-_{ij}$ comprised only of $F_{ij}$'s positive and negative edge weights, respectively. The strength and total weight of $F^\pm_{ij}$ were $s^\pm_i = \Sigma_j F^\pm_{ij}$ and $m^\pm = \frac{1}{2}\Sigma_{ij} F^\pm_{ij}$, respectively. For a given consensus community $\mathcal{C} \in \mathcal{P}(t)$, its positive and negative modularity scores were computed as:

$$q^\pm_\mathcal{C} = \frac{1}{2m^\pm} \sum_{ij \in \mathcal{C}} \left[ F^\pm_{ij} - \frac{s^\pm_i s^\pm_j}{2m^\pm} \right]$$

Large communities, because they consisted of many vertices, also tended to have large modularity scores. To remove this bias, the procedure described above was repeated 5000 times but with vertices randomly assigned to communities $\mathcal{C}_{rand}$, each time resulting in a measurement $q^\pm_{\mathcal{C}_{rand}}$. This score enabled estimates of the expected value ($E[q^\pm_{\mathcal{C}_{rand}}]$) and variance ($\sigma[q^\pm_{\mathcal{C}_{rand}}]$) of each community's modularity score to be obtained. From these estimates, each community's score was standardized and expressed as a z-score:

$$z^\pm_\mathcal{C} = \frac{\left(q^\pm_\mathcal{C} - E[q^\pm_{\mathcal{C}_{rand}}]\right)}{\sigma[q^\pm_{\mathcal{C}_{rand}}]}$$

The sum $z_\mathcal{C} = z_\mathcal{C}^+ + z_\mathcal{C}^-$ was interpreted as an indicator of how well each community $\mathcal{C} \in \mathcal{P}(t)$ mapped onto a functional community. A large, positive $z_\mathcal{C}$ indicated that a community's modularity was much greater than would be expected given its size.

In addition to identifying communities that were more modular than by chance, the scores $z_\mathcal{C}$ were useful for answering several important questions about how stability-derived communities related to rsFC: (i) On average which cortical regions contributed the most standardized modularity; (ii) which pairs of vertices and (iii) which pairs of cortical regions, when assigned to the same community, portended a large standardized modularity score for that community.

From the set of standardized modularity scores, it was straightforward to compute the total contribution of each cortical region by summing the scores of each vertex $v_i$ over all communities in which $v_i$ participated and then aggregating these scores by cortical region.

Another important question was how the co-assignment of groups of vertices or cortical regions to a given community influenced that community's modularity. For example, does assigning vertices $v_i$ and $v_j$ to the same community portend a higher or lower modularity for that community? To identify such groups, a matrix $T_{ij}$ was built and subsequently clustered. Initially, the weights of $T_{ij}$ were set to zero. The weights were updated by considering each community $\mathcal{C}$ and strengthening the connections among all of the vertices assigned to $\mathcal{C}$ by an amount $z_\mathcal{C}$. Thus, the weights of $T_{ij}$ were equal to the sum of standardized modularity over all communities in which vertices $v_i$ and $v_j$ both appeared. An agglomerative, hierarchical clustering algorithm was then used to extract groups of vertices that collectively influenced a community's modularity. This algorithm treated each row in $T_{ij}$ as a feature vector of the matching vertex. Starting with every vertex in its own cluster, clusters were merged over a series of steps until only two clusters remained. At each step, the relationship between every pair of clusters was defined by the average Euclidean distance between the feature vectors of their respective elements. The heuristic for merging clusters was to identify the two clusters whose distance was smallest and to combine their elements, forming a larger joint cluster at the next step. The result of this procedure was a hierarchical tree of related vertices. The tree can be thresholded, revealing a finer or coarser clustering depending on the level of the threshold. At any level, however, these clusters were interpreted as groups of vertices that collectively participated in communities with large standardized modularity scores.

To identify pairs of cortical regions whose co-assignment contributed to a community's having small or large modularity, $T_{ij}$ was down-sampled by aggregating its rows and columns according to cortical region. The elements of the down-sampled matrix contained the sum of modularity accounted for by communities in which each pair of cortical regions' constituent vertices appeared in together.

**Results**

The previous section described a procedure for identifying the connectome's multiscale community structure using the partition stability framework. The association between this structure and observed functional connectivity was assessed by correlating it with rsFC and, separately, by measuring how well consensus communities modularized the rsFC matrix.

Maximizing partition stability at 185 time points logarithmically–spaced over the range $[10^{-3.5}, 10^{+3.5}]$ generated a sequence of partitions for each participant. Each sequence began at the shortest dynamical scale with the partition in which every vertex comprised its own community. A division of the vertex set into two communities characterized the final partition in each sequence, corresponding to the largest dynamical scale. From these partition sequences, a number of statistics were computed at each time point, specifically, the mean and standard deviation partition stability, number of communities, community size, and number of singleton communities. Partition stability and the number of communities declined monotonically while the size of communities increased. The number of singleton communities declined initially, before the maximum number of non-singleton communities reached a maximum value at Markov time $t = 10^{-1.845}$ (Figure 3).

An iterative thresholding/reclustering algorithm was used to generate from the participants' partitions a sequence of consensus partitions, which represented the "backbone" community structure. Examples of consensus partitions at selected Markov times are shown in Figure 4A-F. Each partition represents a division of the cerebral cortex into communities of brain regions poised for information exchange at the corresponding dynamical scale via diffusion dynamics. (For a list of all vertices and their community assignments at each Markov time, see the Supplementary Information).

To compare community structure to brain function, the Pearson correlation of the time-dependent agreement matrix $D_{ij}(t)$ with the empirically-measured rsFC matrix $F_{ij}$ was computed, peaking at a value of $r = 0.50$ at time $t = 10^{0.455}$ (Figure 5). This level of correlation is comparable to earlier studies reporting correspondence between the structural connectome and rsFC, as well as correlations between functional connectivity generated in computational models and empirical rsFC (Honey et al., 2009). An interesting observation was that the peak correlation occurred at a Markov time where there were many communities made up of many vertices ($24.53 \pm 2.06$ communities with an average $20.65 \pm 1.69$ vertices per community) whose stability ($0.85 \pm 0.01$) had not yet begun to decay. Earlier Markov times had slightly greater stability values but were characterized by having far more singleton communities, while partitions at later Markov times had no singleton communities but were marked by extremely low stability values. The consensus partition at this time consisted of fourteen communities, whose topographical arrangement and composition are shown in Figure 6.

As a second means of relating community structure to observed functional connectivity, we measured the extent to which consensus communities were also good functional communities. This process consisted of estimating every consensus community's standardized modularity score. Over a range of Markov times from $t \approx 10^{-1.5}$ to $t \approx 10^{2.5}$, a number of communities had much greater-than-expected modularity (Figure 7A). Mapping these scores onto brain anatomy and summing across Markov times, it was observed that every cortical region contributed positive modularity, though some contributed disproportionately more (Figure 7C). The regions with the greatest modularity (both in terms of peak value and total contribution) were found to be the precentral and postcentral cortex, the lateral occipital cortex, the superior parietal cortex, and the superior frontal cortex. Other regions also had large values, including the rostral middle-frontal cortex, the inferior parietal cortex, as well as the superior temporal, lingual, and fusiform cortex.

The matrix $T_{ij}$ was constructed to measure the pairs of vertices that collectively participated in communities with greater-than-expected modularity (Figure 7B). This matrix was mapped to the space of cortical regions by aggregating the values of vertex pairs that linked cortical regions. This process revealed a number of areas of cortex that, together, participated in communities with large modularity. (Figure 7D). The largest contributing pairs were precentral/postcentral, lateral occipital/fusiform, supra-marginal/postcentral, and rostral-middle-frontal/precentral cortices.

At the vertex level, $T_{ij}$ was clustered to reveal groups of vertices that collectively participated in communities with greater-than-expected modularity (Figure 8A-F). Clustering produced a hierarchical tree, which was cut at different levels to reveal greater or fewer clusters. At a coarse scale, $T_{ij}$ was decomposable into three spatially contiguous modules (Figure 8A): The first module (orange) spanned most of the cortical midline and included superior frontal and paracentral cortices as well as the precuneus; the second module (yellow) was composed of both precentral and postcentral cortices as well as the rostral middle-frontal cortex; the third module (blue) included superior temporal, lateral occipital, lingual, and inferior parietal cortices along with the fusiform area.

Cutting the hierarchical tree at a lower level divided these macro-scale modules into smaller sub-modules. The first module survived largely unchanged for the range of clusters shown in Figure 8, but did fragment slightly as parts of the superior frontal and precentral cortices were split off. The second module was eventually halved into two sub-modules: The first of which was composed primarily of sub-regions of the frontal and precentral cortices; the second module consisted mostly of precentral, postcentral, and supramarginal cortex. The third module underwent a series of splits into four smaller sub-modules (Figure 8D-F): the first sub-module (green) contained primarily lingual gyrus, cuneus, and pericalcarine regions; the second (yellow) was dominated by lateral occipital, fusiform, and superior parietal cortex; the third (cyan) consisted of large portions of temporal cortex, as well as some lateral portions of the orbitofrontal cortex and the insula; and the fourth (blue) was made up of mostly inferior parietal cortex, but also the banks of the superior temporal sulcus along with other areas of the temporal cortex. In principle, $T_{ij}$ could be decomposed further until the number of modules was equal to the number of vertices in the network.

**Discussion**

In this paper we used the partition stability framework to infer multi-scale community structure in the human cerebral cortex. This procedure generated a series of communities over a range of dynamical time scales, beginning with a large number of small communities (fine-scale) and ending with a small number of large communities (coarse-scale). We compared communities to brain function using two approaches: (i) We identified a scale at which the Pearson correlation of an agreement matrix with empirically measured rsFC became maximal, with a peak correlation of approximately $r = 0.50$; (ii) We evaluated the modularity of each consensus community when it was imposed on rsFC, and identified a number of communities that overlapped with rsFC modules. These results suggest that a community's position in both space and time, i.e. where it is located physically/topologically as well as the range of Markov times over which the community appears, provide complementary information for assessing its importance and relevance to network function.

Before discussing these results, it is worth explicitly stating our views of stability maximization and its relationship to the human connectome and communication processes in the brain. Communities derived from maximizing partition stability can be interpreted in multiple ways. The first interpretation regards such communities as being dynamically important – communities reflect the structural properties of a network that bias the trajectory of a random walker exploring the network. If the random walk is taken as a model of message passing or communication among vertices, then stability-derived communities take on even more significance. In such a case, communities represent groups of nodes that more readily exchange messages (communicate) with one another than with the rest of the network. On the other hand is the view that stability maximization, despite its dynamical underpinnings, is simply a useful methodology for identifying community structure across multiple scales; no special functional significance is attributed to communities detected this way. In this discussion, we adopt a view more closely aligned with the first interpretation. Our argument for doing so stems from our conceptualization of the connectome as a communication network – brain regions at different scales signal to one another, passing information from region to region over their anatomical connections. Furthermore, we assert that it is the connection topology that prescribes a region's preference to dynamically link to another region, or for a group of regions to mutually couple. Hence, from our point of view, identifying communities of vertices that are likely to communicate with one another is of great practical interest and can illuminate true functional dependencies.

Our first major finding was that the correlation of the time-dependent agreement matrix, which reflected the community ties between vertices at a given time point, with rsFC reached a clear peak value. This result owed its significance to our interpretation of community structure as a marker of vertices' propensity to dynamically couple with one another. Given this point of view, we can think of the peak correlation as the time point at which the empirical functional couplings became most closely aligned with the communication preference of the structural network as a whole. This result is significant, as the functional coupling of different brain regions has been interpreted as integration of information (van den Heuvel & Hulshoff Pol, 2010). This result is also surprising, as it suggests that an overwhelming proportion of vertices are "tuned" or "poised" for communication at a specific scale.

In terms of the random walk dynamics, this scale is of particular relevance. The time at which the peak correlation occurred coincided with a regime of Markov time characterized by partitions associated with large stability values but also many non-singleton communities. At much earlier Markov times, the average partition stability was slightly greater, but the number of singleton communities vastly outnumbered the non-singleton communities, implying that very few vertices communicate with one another. At later Markov times, stability tended toward zero, so that even though the communities at those times were much fewer in number and made up of more vertices, they were also more ill-defined than those at earlier times. This suggests that the peak correlation occurs within a fairly narrow regime of Markov time in which the number and size of communities strikes a balance with stability, giving rise to a diverse, well-defined repertoire of communities.

The first result, however, only reveals part of the story. Linear correlation is a global measure of correspondence between two variables. When we look at the relationship between community structure and rsFC at a local scale, for instance at the level of individual communities, a different picture emerges. Specifically, we found substantial variability in the timing of and extent to which individual communities contributed to the modularization of the rsFC matrix. Certain communities, even when the partition as a whole was only weakly related to rsFC, had exceptionally large standardized modularity. At the time of the peak correlation, a large number of communities collectively had substantial modularity, which likely contributed to the peak in the correlation coefficient. This result, however, suggests that the relationship between rsFC and community structure is not restricted to a single dynamical scale.

Together these two results paint an interesting picture of a possible communication strategy in the brain. The peak correlation represents a "sweet spot" of communication, a timescale where a large number of vertex pairs are simultaneously poised to take advantage of a diffusion process to communicate information between one another. For any given community, however, the scale at which that community becomes most important likely deviates, even if only slightly, from the time of the peak correlation. This observation prompts the interesting hypothesis that brain dynamics may use a tuning mechanism that grants priority to certain scales of communication while suppressing communication at others. As an example, suppose that in order to execute some cognitive function two regions of the brain need to relay information back and forth. It is not unreasonable to suppose that these regions appear in the same community at some scales but not at others. To facilitate communication between these regions, the tuning mechanism turns from whatever dynamical scale it started in, to the one at which these regions become members of the same community. By moving the regions closer together in this community space, cognitive function is ultimately supported. This sort of selectivity of scale has been observed empirically in the frequency-specific coupling of neuronal oscillators in cortical networks (Buzsáki & Draguhn, 2004). In such networks, oscillators can become coupled over a range of high (short-cycle) to low (long-cycle) frequencies. It was observed that coupling in the high-frequency range tended to recruit oscillators from small, local neuronal pools. In contrast, low-frequency oscillations recruited from and bound together neurons from broader pools and supported the emergence of large-scale, spatially distributed cortical networks. Interestingly, because the time series obtained from any given oscillator can simultaneously have power at both high and low frequency, both short- and long-range coupling is possible. On the surface, these results mirror our own: communities at small Markov times (analogous to high-frequency oscillations) tend to have small diameter and correspond to local, spatially contiguous groups of vertices; communities at long Markov times (analogous to low-frequency oscillations) are more spatially distributed and are comprised of non-adjacent vertices whose relationships are more often indirect.

This tuning hypothesis also has interesting implications related to the specificity of brain regions for cognitive processes. When the brain tunes in to a particular dynamical scale, the communities at that scale tend to trap the flow of information. Thus, each community can be regarded as a complex of vertices which readily exchange information with one another but only sparingly with vertices belonging to other communities. At a given scale, then, it may not be meaningful to refer to an individual vertex's role, but instead refer to the function of the community to which the vertex belongs. This suggests that there may be cases where communication between only small numbers of brain regions is necessary in order to perform some cognitive function, but that the dynamical scale at which these regions communicate necessitates engaging the function of a much larger community. Tuning in to such a dynamical scale would have the desired result of heightened communication between the required regions, but also gives rise to incidental communication between brain regions that are unnecessary for performing the cognitive function in question.

A third interesting finding concerns the relationship of the observed community structure to empirically observed resting state networks (RSN). The communities we observed are generally spatially contiguous, due to the fact that the majority of structural connections identified so far in the human connectome are short-distance and connect nearby regions to one another. Therefore it is unsurprising that we did not observe distributed and spatially non-contiguous RSNs, e.g. the right parietal-frontal and default mode network. However, we do observe highly modular communities in occipital cortex, corresponding quite closely to the primary and extra-striate visual RSNs (lingual gyrus, cuneus, lateral occipital, fusiform). We also observe large modularity in communities involving areas that are typically associated with the somatomotor network (pre- and postcentral cortex). Interestingly, both visual and somatomotor RSNs are generally classified as unimodal networks and thought to comprise tightly coupled and functionally related regions that jointly participate in sensorimotor processes (Bassett et al., 2012; Sepulcre et al., 2012). While the community structures that we observed do not directly correspond to the boundaries of these RSNs, their approximate resemblance is suggestive of a relationship between the dynamic process used to identify communities and their strong functional couplings observed in the course of endogenously driven neural activity. Future work is needed to further clarify the nature of this relationship.

There are a number of limitations of this work that should also be discussed. We exploited a random-walk dynamics to identify vertex communities. Random walks as a simple class of linear dynamics are limited in the types of behaviors they can exhibit. Other classes of dynamical systems, especially non-linear systems, can exhibit far more complex behaviors, including deterministic chaos and parameter sensitivity (bifurcations), among others. We defend our decision to use random walks on the grounds that the partition stability measure depends on this choice of dynamics to analytically define community structure. Furthermore, the well-documented behavior of a random walk drastically simplifies the analysis and interpretation of results.

Even when we restrict ourselves to the random walk class of dynamics, we still have room to complicate our model further with the addition of parameters that bias the random walk, altering the transition preference or the rate at which random walkers leave vertices. These parameters are defined on a vertex-wise basis and afford us means of artificially introducing heterogeneity into the random walk dynamics (Lambiotte et al., 2011). In this article, these parameters were set to fairly conservative values, i.e. a random walker's transition was made without bias and the rate at which transitions took place was equal for all vertices. Note that this parameterization is the most naïve we could have selected – choosing to bias the random walk or to imbue certain vertices with faster or slower transition rates would both require justification.

Lastly, our use of the partition stability framework differed in some aspects from the way in which it has most frequently been applied. Typically, an important step in the framework is to use several robustness tests to verify the measured community structure (Lambiotte, 2010; Delmotte et al., 2011). These tests usually involve measuring the sensitivity of community structure to small changes in edge weight or variations in time – partitions that fail these tests are regarded as being irrelevant. In using the partition stability framework, we chose not to engage in robustness testing. We justify this choice by noting that each participant's connectome can be regarded as a sample drawn from a population of closely related connectome networks. Therefore, obtaining the consensus community structure is an implicit test of robustness. At no point, however, did we test the robustness of community structure as time changed. We argue that establishing the robustness of a whole partition is of little practical interest and that a more revealing test is to identify individual communities that persist over time. We note that our analysis revealed a number of communities that appeared at early Markov times and that remained disjoint for long ranges of time.

**Conclusion**
This article offers novel insight into the relationship between models of dynamical processes unfolding on the structural connectome, community structure, and empirically measured dynamic couplings. By maximizing partition stability, communities at multiple dynamical scales were identified. These communities were then compared to observed patterns of neural activity through a simple correlation measure and also by assessing each community's contribution to the modularization of rsFC. It was observed that a number of anatomical regions contributed disproportionately to this modularization, suggesting that compared to other regions, they might be more likely to engage in a diffusion-like communication process over the timescales when these contributions occurred. Future work is needed to further illuminate the role of different classes of dynamic processes in generating patterns of rsFC in complex brain networks.

**Acknowledgements**
Supported in part by: The National Science Foundation/IGERT Training Program in the Dynamics of Brain-Body-Environment Systems at Indiana University (RB); Swiss National Science Foundation, SNF grant 320030-130090 (AG); Spanish Government grant, contract number E-28-2012-0504681 (JG); Leenards Foundation, Switzerland (PH); JS McDonnell Foundation (OS).

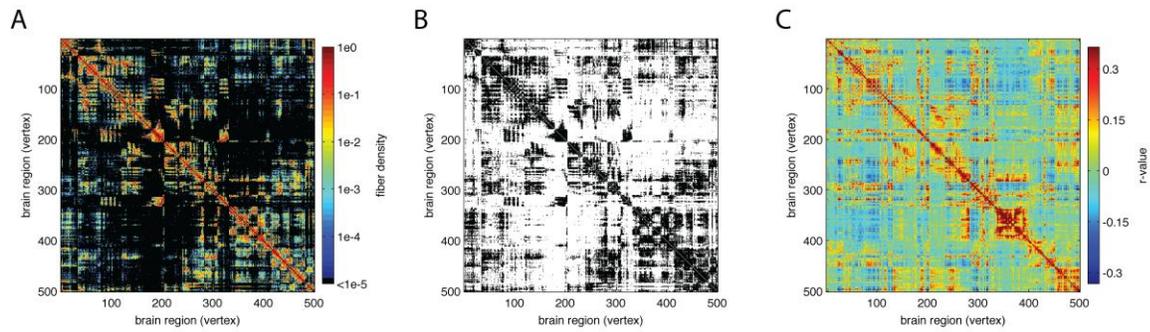

Figure 1. Structural and functional matrices averaged over 40 participants; A) SC weighted: entries indicate the average density of white matter fiber tracts connecting brain regions; B) SC binary: entries indicate the presence {absence} of connecting fibers present in any of the 40 participants; C) rsFC: entries are averaged correlations of the processed BOLD time series during resting state.

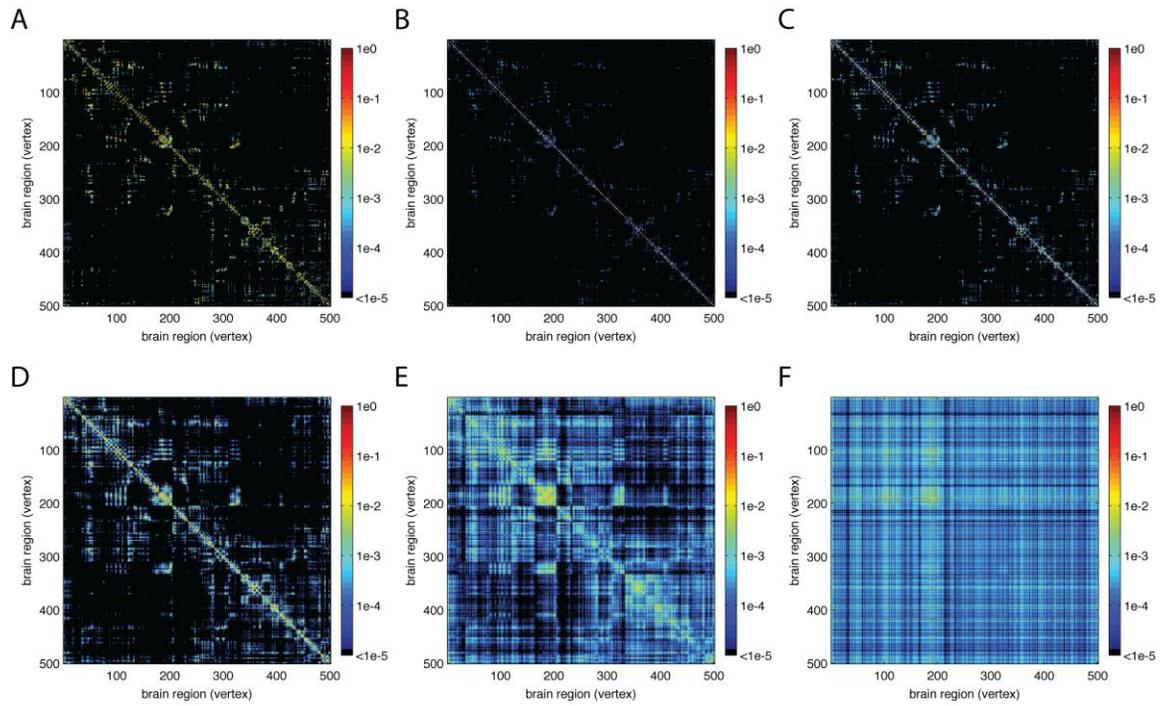

Figure 2. Examples of flow graphs obtained at different Markov times in a single participant; A) Original SC matrix; B-F) flow graphs at times $10^{-2}, 10^{-1}, 10^0, 10^{+1}, 10^{+2}$, respectively, depicting the probabilistic flow of random walkers between pairs of vertices.

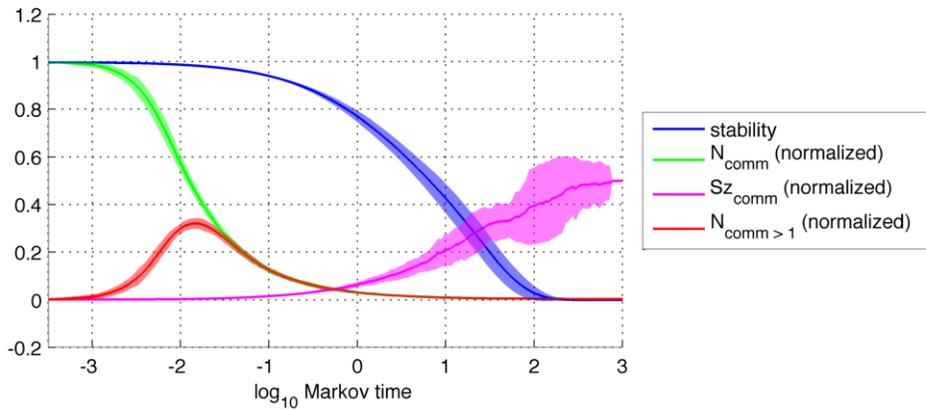

Figure 3. Statistics from community structure obtained from stability optimization procedure (all curves depict mean plus/minus two standard deviations); partition stability (blue); number of communities (green); number of communities with more than one vertex (red); average community size (magenta). Community size, number of singleton communities, and number of communities are normalized by size of the network (501 cortical regions in the right hemisphere) so that they scale between 0 and 1.

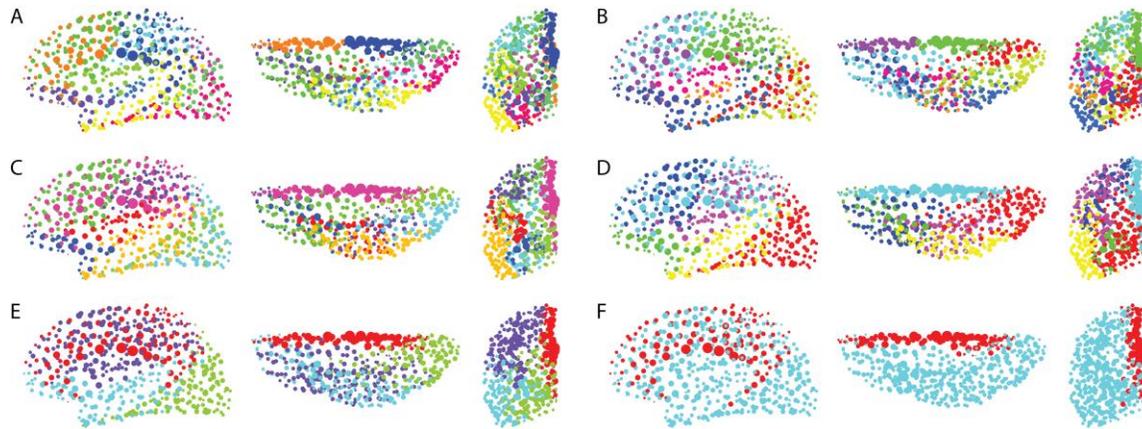

Figure 4. Consensus partitions obtained at different Markov times; A) $t \approx 10^{-0.23}$ with twelve communities; B) $t \approx 10^{0.08}$ with ten communities; C) $t \approx 10^{0.273}$ with eight communities; D) $t \approx 10^{0.53}$ with six communities; E) $t \approx 10^{0.76}$ with four communities; F) $t \approx 10^{2.89}$ with two communities. Vertex size is proportional to strength.

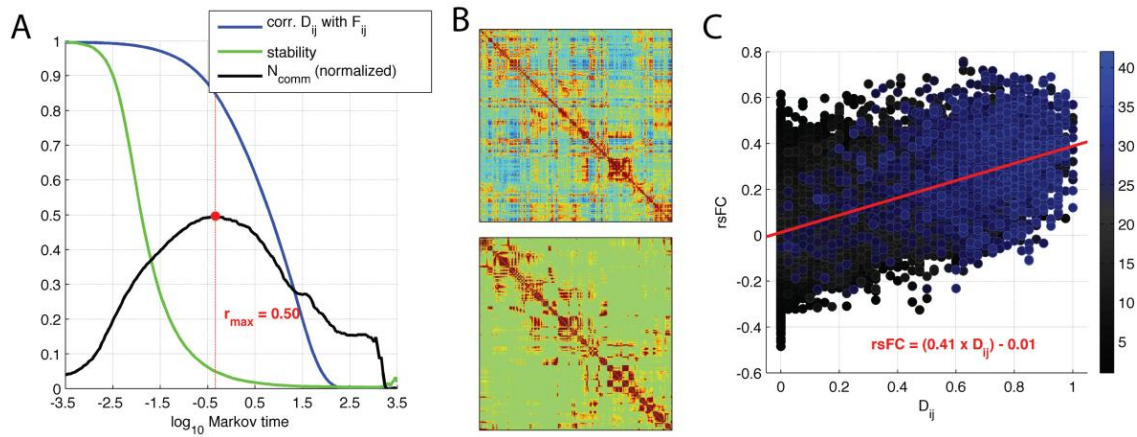

Figure 5. Summary of correlation between agreement matrix $D_{ij}(t)$ and resting-state functional connectivity (rsFC); A) Pearson's correlation value (black) at each time point. Superimposed on this plot are the mean stability (blue) and mean number of community (green) curves. The time at which the max correlation value occurs falls within a narrow regime where the number of communities and stability are both substantial; B) $D_{ij}(t)$ at peak correlation value (top) and rsFC (bottom); C) Scatter plot of $D_{ij}(t)$ and rsFC at time of peak correlation (linear fit - red line). The color of each "o" denotes the number of participants in which the corresponding structural connection was also present.

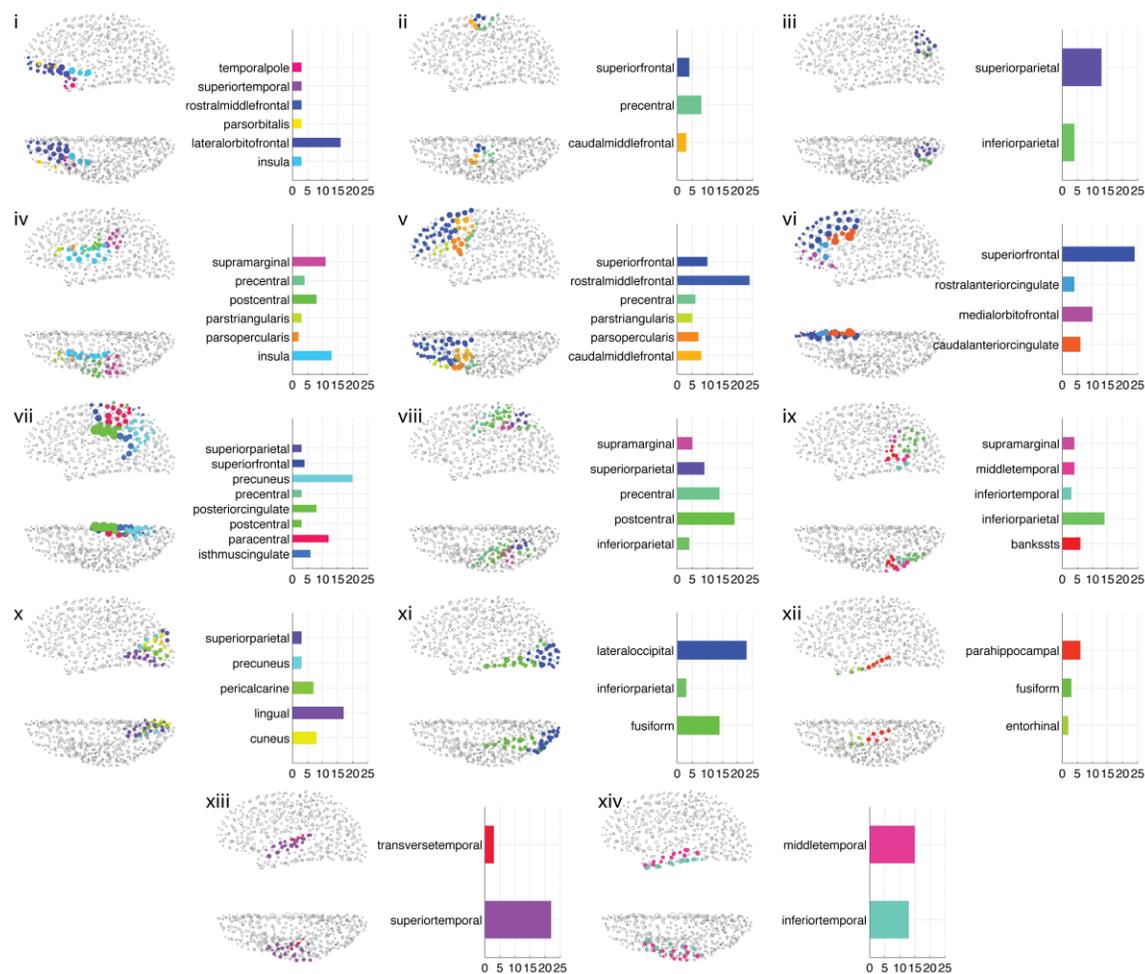

Figure 6. Summary of 14 consensus communities at the time of peak correlation. In each plot, vertices are colored gray if they do not belong to the corresponding community. The color of any non-gray vertex indicates the cortical region that that vertex belongs to. Once again, vertex size is proportional to strength.

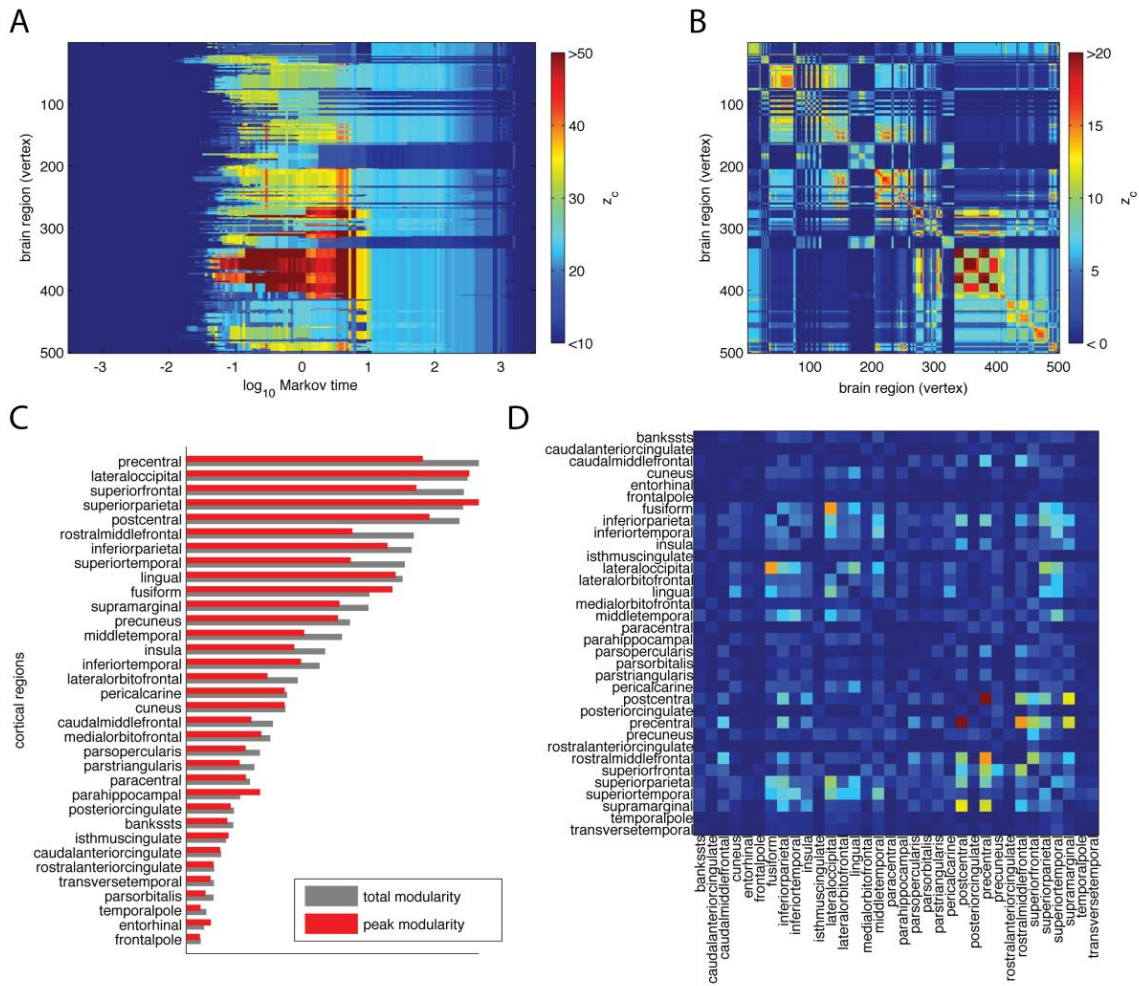

Figure 7. Summary of modularity scores of communities at different times; A) sum of positive and negative normalized scores by vertex over time; B) weighted matrix where each community co-assignment was weighted by its score; C) anatomical region-by-region average and peak standardized modularity score; D) weighted agreement matrix aggregated across by anatomical region.

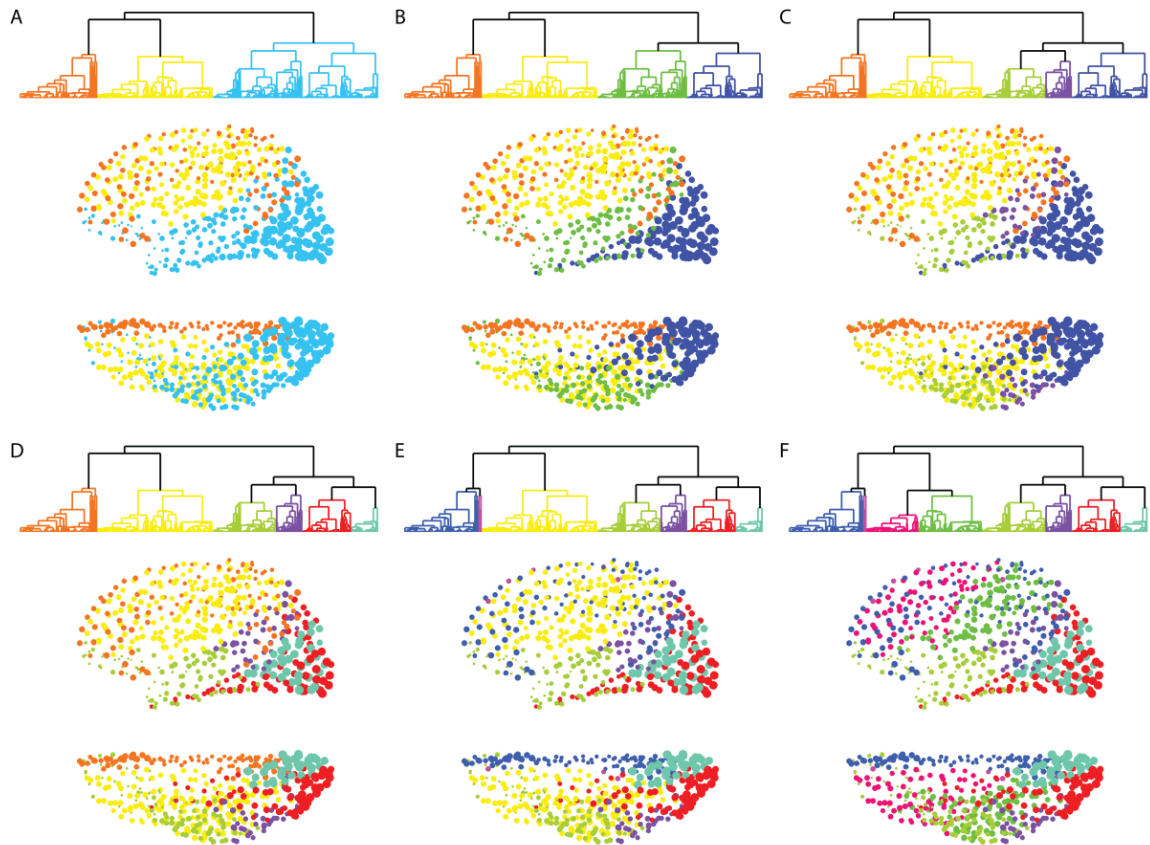

Figure 8. Summary of hierarchical tree cut at different levels; A-F) Cuts reveal spatially contiguous maps of three to eight clusters.